# Fast and Adjustable Resolution Grazing Incidence x-ray Liquid Surfaces Diffraction achieved through 2D Detector

***Philippe FONTAINE, Michel GOLDMANN***

***Michel BORDESSOULE, Alain JUCHA***

*Laboratoire pour l'Utilisation du Rayonnement Electromagnétique (LURE, UMR CNRS 130), Centre Universitaire Paris-Sud, Bât. 209D, BP 34, F-91898 ORSAY CEDEX, France*

Running Title : Liquid Surfaces GIXD through 2D Detector

**Abstract**

We developed a setup using a two dimensional camera for Grazing Incidence x-ray Diffraction (GIXD) on Langmuir monolayers and more generally for surface diffraction on two dimensional powders. Compared to the classical setup using a linear detector combined with Soller's slits, the acquisition time is reduced of a factor of at least 10 (from more than one hour to a few minutes) using the same x-ray source (synchrotron bending magnet) with a comparable signal to noise ratio. Moreover, using an horizontal gap slit, the experimental resolution can be adjusted and for small values of the gap, better resolution can be achieved compared to the one obtained with the linear detector.

## 1- Introduction :

Grazing Incidence x-ray Diffraction (GIXD), since its first use in 1981[1], has enabled major breakthroughs in surface structures investigation. The study of fresh solid single crystal surfaces by GIXD revealed complex surface reconstruction phenomena [1, 2]. This method has been rapidly applied in soft condensed matter to study “soft surfaces” (Langmuir, Langmuir Blodgett, self-assembled films). Indeed, following the first application of GIXD simultaneously by Dutta & al[3] and Kjaer & al [4] on Langmuir monolayers, numbers of similar experiments have been performed by various teams [5-8] revealing the rich polymorphism of these films and leading to the proposition of a generic phase diagrams [9, 10], a breakthrough in this field. It demonstrates the major interest of GIXD experiments on soft interfaces which appears as complementary to surface pressure measurement [11], optical microscopy (Brewster Angle Microscopy [12, 13], epifluorescence microscopy [14]) and x-ray reflectivity [15] measurements. More recently, GIXD revealed the complex behavior of the compressibility of fatty acid Langmuir monolayers [16].

Before the use of GIXD, the lack of molecular organization information has precluded the use of Langmuir monolayers. However, they have many applications in fundamental and applied science. They are considered as two dimensional system to study phase transitions and test numerical simulations [17, 18]. Their studies give valuable information about interactions between biological molecules dissolved in the aqueous subphase (drugs, proteins, peptides, ions …) and the phospholipid monolayer [19-21]. The air/water interface can also be used to obtain new materials. 2D polymers can be synthesized within or below Langmuir monolayers [22, 23] and recently, metallic layers have been produced below Langmuir films by different synthesis methods such as radiolysis[24] or electrochemistry[25].

The current trend of Langmuir monolayers research leads to high resolution setup for line shape analysis of diffraction peaks on one hand. Indeed, the widespread setup to perform GIXD on Langmuir monolayers uses a one dimensional Position Sensitive Detectors (PSD) with a Soller collimator which defines the setup to an average in plane resolution, about $0.07nm^{-1}$ for the best available set of slits. On the other hand fast acquisition setup for kinetics measurement in chemistry or biology are needed. In these studies, small changes of the molecular arrangement in the monolayers have to be recorded which needs not only good resolution but also high statistics and thus large counting time. In some other cases, studies concern fast adsorption kinetics (less than one hour) which needs a fast acquisition setup. However, improving the x-ray flux does not always represent a valuable alternative, since the x-ray beam could damage the sample. Thus, efforts should be made on the acquisition setup.

Fast acquisition setup has been proposed using a strongly focused incident beam (spot size of typically hundred of microns). With such incident beam, the sample is considered as a point and a two dimensional detector (image plate) records the diffraction spectra [26, 27]. Although short acquisition time can be achieved using such a setup, the resolution is poor ($0.27nm^{-1}$) [26] and the high flux on the sample could be a limitation in some cases. In this paper we propose a new experimental setup based on the use of a 2D camera which increases the possibility of GIXD without changing the incident beam geometry. Indeed, we show that this setup allows faster acquisition. Moreover, the experimental resolution can be tuned by simply adjusting a slit gap. This paper is divided in three parts. In the first one, the principle of GIXD and the classical setup are briefly presented. In the next part, the new setup based on the 2D detector is exposed. Finally some results testing the new ability of this setup are given and discussed.

## 2- GIXD principle and classical experimental setup :

The GIXD setup to measure surface structure is based on the following principles. To avoid scattering from the substrate and improve the signal to noise ratio, the incident beam reaches the interface at grazing incidence, below the critical angle of the air/substrate interface. This can be obtained by considering the properties of the respective refraction index for x-ray of the two medium forming the interface (liquid – gas, liquid – vacuum, solid – gas, solid – vacuum, …). For a gas or vacuum this index is one. For x-ray, the refraction index of liquid or solid is $1-\delta+i\beta$, $\delta$ ranging from $10^{-6}$ to $10^{-5}$ and the absorption coefficient $\beta$ ranging from $10^{-11}$ to $10^{-10}$ [6, 28]. Then, considering the Fresnel's law at such interface when the incoming x-ray beam propagates in the gas or vacuum medium, total external reflection may occurs. For incidence angles (measured between the interface plane and the incident beam) below the critical angle of total external reflection $\alpha_c$, the incident x-ray beam is quasi-totally reflected by the interface as no transmitted wave propagates in the condensed phase. However, an evanescent wave propagates in this phase along the interface plane as its intensity exponentially decreases with the distance to the interface. This evanescent wave is used as a source to probe the interface structure.

The critical angle of total external reflection $\alpha_c$ is given by :

$$\cos(\alpha_c)=1-\delta\ .$$

For angle of incidence $\alpha_i$ below $\alpha_c$, the penetration depth of the evanescent wave is given by:

$$\xi=\frac{\lambda}{4\pi}\frac{1}{Re\left(\sqrt{\cos^2(\alpha_i)-((1-\delta)+i\beta)^2}\right)}.$$

For the air-water interface, at a wavelength $\lambda = 0.160 nm$, one obtains $\delta = 3.85 \cdot 10^{-6}$, leading to a critical angle of $2.77 mrad$. Then, for an incidence angle $\alpha_i = 2.35 mrad$ the penetration depth is $\xi = 4.6 nm$. Due to this small incidence angle, a small divergence in the plane of incidence is mandatory and leads to the use of synchrotron sources.

Considering the beam thickness, one obtains a few centimeters length footprint on the sample for such incidence angles. All molecules located in this footprint are scattering centers. Moreover, Langmuir monolayer are usually 2D powders (all domains are lying in the plane of the interface, but the 2D crystals are randomly oriented in the plane) and considering the sample-detector distance (≤ 1m), collimation of the scattered photons is mandatory. Organic molecules are mainly composed of "light" atoms whose scattering cross sections are rather weak. Then the use of Soller's collimator have been a worthwhile solution to improve the measured intensity. Indeed, it allows to enlarge the observed area almost proportionally to the in plane dimension of the collimator.

The first experiments dedicated to GIXD on Langmuir monolayer were using a point detector after the collimator to collect the scattered photons. However in plane information was only recorded in an horizontal scan. To determine the molecular orientation with respect to the 2D lattice (tilt of aliphatic chains), one must perform vertical scans (along the $Q_z$ direction) on the diffraction peaks in order to obtain the shape of the diffraction rods (due to the 2D character of the sample)[2,6]. A first improvement was to replace this detector by a vertical one dimensional Position Sensitive Detector (PSD) mounted on a horizontal 2-theta circle. In-plane scan of the $2\theta$ angle allows then to record simultaneously the "integrated" $Q_{xy}$ scan and the vertical intensity distribution. It appears that the efficiency of this setup is mainly limited by the Soller's collimator performances. Indeed, the in plane resolution is defined by the corresponding slits aperture. Moreover, some scattered photons are not detected. These photons are scattered by molecules not located in the area defined by the interception of the x-ray footprint and the Soller's collimator projection. To maximize the measured intensity at a $Q_{xy}$ (or $2\theta$) position, one expects the larger in plane collimator as possible (a maximum of parallel and identical slits) but also the larger size along the vertical direction (to collect the larger $Q_z$ range). The best available collimators are provided by *JJ X-ray* in Denmark. Its aperture is $2.8mrad$ with about $20mm$ width and $100mm$ height. The transmission is about 50%.

The in plane resolution can be strongly improved by replacing the Soller's slit collimator by a single or double crystal analyzer[29]. One can reach in this case resolution as good as $0.008nm^{-1}$. However, the measured intensity is decreased of an order of magnitude. Thus, the use of 3$^{rd}$ generation synchrotron source is mandatory. Moreover, the principle of the setup remains identical to the one with Soller's slit.

Figure 1 represents the experimental setup dedicated to liquid surface diffraction on the D41 beam line of the DCI storage ring at LURE (Orsay, France). The x-ray source is a bending magnet. The beam is monochromatized at $\lambda = 0.1605nm$ ($7.7KeV$) by a (111) Germanium plate $17cm$ long, with an asymmetrical cut angle of $9.88°$. This plate is bent following a classical procedure previously described [30] in order to provide horizontal focussing. The focus spot is located on an x-ray camera which is $3.68m$ downstream the monochromator ($0.98m$ downstream the sample). Its horizontal size is estimated to $3mm$. The beam is then collimated by a set of narrow horizontal and vertical slits. Due to the horizontal nature of the liquid-gas interface, the synchrotron x-ray beam must be deflected downward to impinge on the liquid surface. This is achieved by a flat mirror ($150mm$ long) whose inclination, precisely controlled, defines the angle of incidence.

The air scattering is reduced by enclosing the Langmuir trough in a gas-tight box flushed with Helium gas and equipped with Kapton windows. The intensity of the incident beam ($I_0$) is measured by a NaI detector monitoring the scattered intensity of the direct beam by a Kapton foil. The diffracted beam ($I_c$) is measured with a $5cm$ depth vertical position sensitive detector (PSD) filled with $Ar/CO_2$ gas mixture ($5\% \, CO_2$) at 1.65 bar. This depth insures an efficiency of almost 85% at $7.7keV$. The Soller slit collimator is positioned in front of the PSD inside a gas-tight box flushed with Helium gas. Its acceptance corresponds to a scattering wave vector resolution of $0.07nm^{-1}$ at $Q_{xy} = 15nm^{-1}$ . The out of plane signal is measured at each in-plane wave vector ($Q_{xy}$) by the PSD. The exit angle angular range is $0-13°$ corresponding to a $Q_z$ range from $0$ to $8nm^{-1}$.

## 3- GIXD setup using a 2D detector

### 3-1 Principle

The goal of a GIXD setup is to collect the scattered photons from the larger part of the illuminated area. The new setup is based on two interesting features. As mentioned above, the grazing angle of incidence (typ. $2mrad$) leads to a wide illuminated area. For example, $200\mu m$ vertical thickness of the incident beam under $2mrad$ of incidence leads to a $100mm$ longitudinal size footprint. The lateral dimension of the footprint (transverse size of the incident beam) is about $2mm$. It will be neglected in the following. The other feature is the 2D powder nature of these monolayers. The 2D powder nature of layer implies that we do not need to rotate the sample to align crystal with the beam. Within the footprint, one always obtains crystallites correctly oriented for any $Q_{11}$ wave vector. Moreover, the $Q_{xy}$ integrated in plane scans give information about the correlation length of the molecular organisation. If the Full Width at Half Maximum (FWHM) is larger than the experimental apparatus resolution ($\Delta Q_{xy}$), this FWHM is inversely proportional to the coherence length after deconvolution of the spectra. Otherwise, if the FWHM is equal to $\Delta Q_{xy}$, the diffraction peak is resolution limited. The correlation length is thus larger than $1/\Delta Q_{xy}$.

In this new setup, we replace the PSD and the Soller's slits by a two dimensional detector and a single thin vertical slit between the sample and the detector, as represented in figure 2. In this configuration each pixel $M$ of the two dimensional detector measures the scattered intensity from a given point of the illuminated area under a $2\theta(M)$ angle, selected by the slit. This angle depends on the position of the $M(u,v)$ pixel on the detector, on the sample-slit distance $\ell$ and on the sample-detector distance $L$. As the sample is a real 2D powder, a large diffraction spectrum can be measured simultaneously (without "scanning" the reciprocal space) by only adjusting the respective value of $L$, $\ell$ and the slit aperture. In Appendix 1, the $2\theta$ angle is analytically computed as a function of $u$, the horizontal position of the pixel on the detector, and of the lengths $L$ and $\ell$. For example, using a 2D detector of horizontal size $D = 92mm$, with $L = 595mm$ and $\ell = 215mm$, the in-plane $Q_{xy}$-range measured by the detector, is $10 - 20nm^{-1}$ for $2\theta_0 = 22°$, the angle between the direction of the direct beam and the arm bearing the 2D detector, and $\lambda = 0.160nm$. Of course, the $Q_z$ intensity distribution is simultaneously recorded and deduced from the vertical coordinates ($v$) on the detector.

The resolution of the setup is mainly governed by the horizontal gap of the slit $g$ associated to the $L$ and $\ell$ distances as presented on figure 3. In the Appendix 2, the $\Delta 2\theta(u)$ angle accepted at the point $M(u)$ of the detector and thus the $\Delta Q_{xy}$ resolution is analytically determined for a parallel beam (no divergence). Figure 4 shows the theoretical evolution of $\Delta 2\theta$ along the horizontal axis of the detector ($u$) for a $0.7mm$ horizontal gap, and for similar distances ($D = 92mm$, $L = 595mm$, $\ell = 215mm$, $2\theta_0 = 17.735°$). This calculation demonstrates that $\Delta Q_{xy}$ evolves along the horizontal axis of the detector. The minimum (better resolution) is reached at the edges of the detector, the maximum (worst resolution) is close to the middle of the detector. This is due to the asymmetric position of the detector with the incident beam axis. However, this variation remains small, less than 2% in this case. Thus, we will always refer to the maximum value for calculated $\Delta Q_{xy}$ in the following. For a vertical slit opened at $0.7mm$, $\Delta Q_{xy}$ is $0.072nm^{-1}$, which is the resolution of the "classical" PSD and Soller's slit setup. Since smaller gap can be reached or distances increased, the resolution of this new setup can be improved as far as the in plane resolution of the 2D detector is not reached. In our case, the horizontal resolution ($\delta u$) is $200\mu m$. By calculating the error $\delta Q_{xy}$ from the calculated $Q_{xy}$ formulae obtained in appendix A-1, one obtain a minimum resolution for the setup of $0.02nm^{-1}$ which is much better than the resolution of the 1D-PSD setup.

The resolution can be easily adjusted by just varying the setup's parameter, as shown in figure 5. The variation of slit gap between 0 to 5 mm varies the resolution from 0 to $0.5nm^{-1}$ (figure 5-A); varying $\ell$, the sample-slit distance, between 100 to 300 mm adjusts the resolution from $0.05nm^{-1}$ to $0.09nm^{-1}$ (figure 5-B) for $g = 0.7mm$; the variation of $L$, the sample-detector distance from 1000 to 300 mm adjusts the resolution from $0.04nm^{-1}$ to $0.3nm^{-1}$ for $g = 0.7mm$ (figure 5-C). Of course, varying these three parameters allows to adjust the resolution, the flux and the $Q_{xy}$ range. One can then easily transform the setup from a low resolution (or large $Q_{xy}$ range) high flux experiment into a higher resolution (small $Q_{xy}$ range) lower flux experiment.

### 3-2 Experimental setup

The vertical slit is made from a silicon single crystal (Si 111 wafer, 2 inches diameter, $1mm$ thick) cut in two parts. The use of a single crystal avoids diffuse scattering or diffraction of the diffracted beam by the slit edges. The two parts of the silicon wafer are fixed to a slit holder which have two degrees of freedom: a global translation to center the slit on the line between the center of the 2D detector and the center of the goniometer, and an aperture of the horizontal gap. Horizontally, the detector is centered on the axis of the 2-theta arm. The bottom of the detector window is adjusted vertically with the level of the water surface.

The home built 2D detector is a gas-filled (xenon 85%, ethane 15%) wire detector. The $100mm \times 100mm$ cathode plane is segmented in $32 \times 32 = 1024$ squared pads. As the required vertical resolution could be low, the anode plane is made of 32 horizontal 20µm diameter wires, with the same pitch as the cathode pads. The wires are spaced each 3.17mm, thus allowing a relatively low operating voltage of 1800Volt. This wire plane is located at the center of the space between the beryllium window and the cathode plane. The effective size of the entrance beryllium window is $92 \times 92mm$. Its thickness ($0.5mm$) enables to seal the detector and fill it with xenon-ethane gas mixture at a pressure of $1.1bar$. As the mean expected rate was around $10^3$ counts per second, a relatively slow but sensitive, low noise electronic was chosen, namely the CERN Gassiplex chip[31]. A 1D detector using the same electronic has been described elsewhere [32]. The electronic system, located within the detector, is connected to a PC computer by a fast digital link. The 32 horizontal active anode wires are connected to a trigger. This trigger rejects or counts events on a dead-time criteria, or on an amplitude criteria. A first kind of rejected events are the ones separated by less than the measured 65µs fixed[33] dead-time of the detector. This dead-time is the delay needed by the 12bits-1µs Analog to Digital Converter to digitize each accepted event. As the transfer and computational burden of the computer is light, the apparent dead-time is the one of the detector. A second kind of rejected events are those who do not comply with the two thresholds of a window discriminator. Thirdly, even once an event has been accepted by the trigger and then transferred to the computer, it can be rejected for various reasons by the software, mainly because being too near to the edges of the cathode or because of an incorrect charge distribution. As the 500ns pulse-pair resolution of the trigger is quite low compared to the average time between the events, of the order of 1ms, the probability of accepting nearly simultaneous events is low. As the 2D histogram is a computational result, the number of pixels on each axis is free. The efficiency is determined by the $0.5mm$ thickness of the beryllium window and the $7.5mm$ of gas. The dynamics of the detector is better than $10^4$ count per second. The spatial resolution (FWHM) of the detector is about $200\mu m$ along the wire (delay lines) and $500\mu m$ perpendicularly, using a center of mass calculation of the charge distributed on the adjacent wires. The number of pixels have been set to 1024 horizontal $\times 256$ vertical. The recorded noise on the whole detector in absence of x-ray illumination is about 3 count per second.

Since the best resolution is obtained along the wire, their horizontal location enables a good in-plane resolution which is mandatory for GIXD measurement. Vertically, the resolution is thus lower but sufficient since interface diffraction peaks exhibit smooth variations of the intensity along the vertical rods in most cases.

The space between the detector window and the slit is enclosed in a gas tight box flushed with helium in order to reduce scattering and absorption by the air. Figure 6 is a picture of the whole setup as it was built on the D41 beam line at LURE.

## 4- Results and Discussion

Tests on this new setup have been performed on two different systems: a well-known phospholipid to compare acquisition with the 1D and 2D detection, and a fluorinated fatty acid to test the resolution of the 2D setup.

The phospholipid was DPPE (L-α-Dipalmitoylphosphatidylethanolamine). The molecules were purchased from Sigma with a purity better than 99% and used as received. The molecules are dissolved in mixtures of chloroform and methanol (9 :1) (Fisher, certified HPLC) in order to obtain a spreading solution of $1 mmol.\ell^{-1}$. The fluorinated fatty acid, $CF_3-(CF_2)_{10}-COOH$, was purchased from Sigma with a purity better than 95% and used as received. The molecules are dissolved in mixtures of n-hexane and ethanol (9:1). Surface tension was measured by the Wilhelmy plate method[11]. The plate is made using filter paper $2mm$ large and $0.1mm$ thick. It hangs to a surface pressure sensor (Riegler & Kirstein Gmbh, Wiesbaden, Germany). The accuracy of the measurement was better than $0.1 mN.m^{-1}$. All experiments are performed at $19°C$.

### 4-1- Diffraction tests and comparison with the 1D setup

Figure 7 shows diffraction spectra of a DPPE Langmuir monolayer compressed at $40mN.m^{-1}$ measured with the two kinds of setup: the line represents the spectrum obtained with the 2D detector's setup ($D = 92mm$, $\ell = 215$, $L = 595mm$, $2\theta = 20.245°$, $g = 1mm$), the points represent the spectrum measured with the classical 1D-PSD setup previously described. The intensity was normalized to the peak maximum. The two spectra exhibit the same diffraction peak at $15.14 nm^{-1}$ as expected from the literature[34, 35]. It demonstrates the ability of the two dimensional detector's setup to measure diffraction spectrum. However the time for the acquisition are quite different. It takes one hour to scan over the $2\theta$ angle the 1D detector (1 min per point) and 3 minutes (depending on the statistics needed) to measure the spectrum with the 2D detector. The signal to noise ratios are similar for the two setups, estimated to 1.7 for the 2D detector and to 2.5 for the 1D detector. Notice that the sampling is 3 times lower for the 1D setup.

The vertical distribution of the diffracted intensity measured by the new setup is presented on figure 8 where the $Q_{xy} - Q_z$ contour plot of a DPPE monolayer compressed at 8 different surface pressures are depicted showing from figure 8-A to figure 8-F the decrease of the tilt angle of the hydrocarbon chains in the $L_2$ like phase (Next Neighbour tilt) and between figure 8-F and 8-H the transition from the tilted $L_2$ phase to the LS untitled phase. The acquisition of each image has taken 5 minutes. On figure 9 is presented the same evolution of the diffraction pattern but measured with the 1D-PSD setup: decrease of the tilt angle from figure 9-A to figure 9-C and after the transition to the LS phase in figure 9-D. Acquisition time is about 120min (2 hours) for each spectrum. Comparison between figure 8 and 9 shows that the two setups lead to the determination of the same parameters (in plane and out of the plane peak position). The overall shape of the pattern are equivalent with a weaker statistic for the 2D setup compared to the 1D setup. However, the acquisition time of the 2D setup can be easily increased and thus the statistics. Thus the determination of Langmuir monolayers phase diagram can be achieved quickly and accurately using the new setup.

### 4-2- Resolution of the setup vs. slit gap

In order to test the resolution, we have studied Langmuir monolayers of a perfluorinated fatty acid $CF_3-(CF_2)_{10}-COOH$, which is known to exhibit resolution limited diffraction peaks with the "classical" GIXD setup[17]. In this case, the shape of the diffraction peaks is determined by the resolution function of the experiment.

Figure 10-A & B present the in-plane, integrated spectrum and $Q_{xy}-Q_z$ contour plot respectively, of the perfluorinated fatty acid monolayer at surface pressure of $20mN.m^{-1}$ recorded with the 2D detector setup with slit gap $g=0.25mm$, $L=595mm$, $\ell=240mm$ and $2\theta_0=17°$. It exhibits a single diffraction peak as expected corresponding to a perfect hexagonal arrangement of the fluorinated chains. The peak position is $12.5141\pm0.0006nm^{-1}$ in agreement with previous measurements [17].

Figure 10-C gives the "real" Full Width at Half Maximum (FWHM) determined by taking the difference between the Q value corresponding to half maximum intensity (not result of fit) of diffraction peaks recorded with the 2D detector for different values of the slip gap ranging from $3.7mm$ to $0.25mm$. Decreasing the slit gap leads to the decrease of the width of the peak. No evolution of the peak position is observed (not shown). To compare with the theoretical calculation of the resolution, figure 10-C presents the result of the calculation of appendix 2 (thin continuous line). Although the order of magnitude of the measured and calculated FWHM are correct, the evolution with slit gap is not completely described by the calculation. For large and small slit gap values, the experimental FWHM saturates. At intermediate slit gap, the evolution of the FWHM is linear but do not agree to the calculated one. In order to improve the description and the comprehension of the resolution of the experiment, we performed more detailed calculation taking into account the real optical properties of the beam, namely the convergence $\alpha_0$ of the incident beam due to focalisation, the size of the footprint due to the size of the beam, the gaussian distribution of the intensity

within the beam. Details of the calculation are given in appendix 3. The best calculated curve is the thick continuous line in figure 10-C. The parameters used for the calculation are given in Table 1. The fixed parameters are the length $L$, $\ell$, the size of the detector $D$, the $2\theta_0$ angle, the longitudinal size of the footprint $E$ which is fixed by the mirror size, and the wavelength. Their values are the one measured on the experiment. The only adjusted parameters are the convergence of the beam which cannot be easily measured on our setup but estimated to a few milliradians, and the transverse ($w_y$) and longitudinal ($w_x$) FWHM of the intensity gaussian distribution. These parameters have been varied to study their effect on the calculated curves. They are depicted by the lines of figure 10-D. The transverse FWHM ($w_y$) of the incident beam has no significant influence on the in-plane resolution since curves with $w_y = 0.5mm$, $2mm$ (see thick line of figure 10-C) and $10mm$, are identical. However the longitudinal FWHM of the intensity distribution $w_x$ has a significant effect on the curve as shown on figure 10-D by the calculation with $w_y = 75mm$. Its value controls the saturation of the FWHM of diffraction peaks for the large slit gap values. The longitudinal intensity distribution influences the shape of the diffraction peaks since the points corresponds to intensity scattered by regions of the footprint far from the center of the footprint where the incident intensity is damped by the gaussian nature of the beam. Finally, the beam convergence $\alpha_0$ controls the saturation of the experimental FWHM of the diffraction peaks for small slit's gap values as shown in figure 10-D by the calculation with $\alpha_0 = 0$. The convergence limits the resolution of the apparatus since the incident in plane wave vector $\mathbf{k}_i$ exhibits a distribution of angle which limits the resolution. The use of a parallel beam will further improve the resolution.

Using a two dimensional detector combined with a single vertical slit, we achieve the measurement of the diffraction pattern of Langmuir monolayers with an equivalent in-plane resolution and a faster acquisition time than the usual setup based on Soller slits and 1D Position Sensitive Detector. Moreover, resolution versus flux or $Q_{xy}$ range can be easily adjusted with this new setup, depending of the need of the measurement (kinetics, structure determination etc…). Finally, better resolution can be reached by improving the incident beam properties (divergence) and the distances.

**Acknowledgement :**

The authors thanks S. Megtert for assistance in the development of the 2D detector.

**APPENDIX :**

A-1- **Scattering wave vector determination:**

In figure 2, we call $L$ the sample to detector distance ($L = OO'$), $\ell$ the sample to vertical slit distance ($\ell = OF$), and N the point of the incident beam axis seen by the point M of the detector of horizontal coordinate $u = O'M$. In the $(ONF)$ triangle, the summation over the three angles should be equal to $\pi$, *i.e.*:

$$2\theta_0 + (\pi - 2\theta(u)) + \alpha = \pi$$

with $\alpha$ the angle between $OO'$ and $NM$. This angle can be determined using the $(O'MF)$ triangle :

$$\tan\alpha = \frac{u}{L-\ell}$$

Thus the diffraction angle is given by:

$$2\theta(u) = 2\theta_0 + Arc\tan\left(\frac{u}{L-\ell}\right)$$

Finally, the in-plane scattering wave vector $Q_{xy}$ measured at a distance $u$ from the center of the detector is given by :

$$Q_{xy}(u) = \frac{4\pi}{\lambda}\sin\left(\frac{2\theta(u)}{2}\right) = \frac{4\pi}{\lambda}\sin\left(\frac{1}{2}\left(2\theta_0 + Arc\tan\left(\frac{u}{L-\ell}\right)\right)\right)$$

**A-2- Determination of the theoretical resolution of the apparatus for a parallel beam:**

A visualization of the $\Delta 2\theta(u) = 2\theta_2 - 2\theta_1$ angle which is seen by the point $M(u)$ on the detector through the vertical slits is given on figure 3.

In the ($Ox, Oy$) coordinates system given in figure 3, the coordinates of the $M(u)$ point are:

$$M = \begin{pmatrix} x_M = L\cos(2\theta_0) - u\sin(2\theta_0) \\ y_M = L\sin(2\theta_0) + u\cos(2\theta_0) \end{pmatrix}$$

The coordinates of the $F_i$ points which are the edges of the vertical slit as represented on figure 3, are given by :

$$F_i = \begin{pmatrix} x_{F_i} = \ell\cos(2\theta_0) + (-1)^i \frac{g}{2}\sin(2\theta_0) \\ y_{F_i} = \ell\sin(2\theta_0) - (-1)^i \frac{g}{2}\cos(2\theta_0) \end{pmatrix}$$

Since the angle $2\theta_i$ is also found between $F_iM$ and the horizontal axis, this angle is given by

$$\tan(2\theta_i) = \frac{y_M - y_{F_i}}{x_M - x_{F_i}} = \frac{(L-\ell)\sin(2\theta_0) + \left(u + (-1)^i \frac{g}{2}\right)\cos(2\theta_0)}{(L-\ell)\cos(2\theta_0) - \left(u + (-1)^i \frac{g}{2}\right)\sin((2\theta_0)}$$

Finally the scattering wave vector resolution is given by :

$$\Delta Q(u) = Q_2 - Q_1 = \frac{4\pi}{\lambda}\left(\sin\left(\frac{2\theta_2}{2}\right) - \sin\left(\frac{2\theta_1}{2}\right)\right)$$

**A-3- Theoretical resolution of the apparatus for a convergent incident beam**

The total intensity measured at point $M(u)$ on the 2D detector is the result of the double integration over $\theta$ from $2\theta_1$ to $2\theta_2$ and over the $s$-coordinate along the line coming from M with an angle of $\theta$ with the incident beam direction as shown in figure 11:

$$I(u) = \int_{2\theta_1}^{2\theta_2} d\theta \int_{-\infty}^{+\infty} ds I_0(s,\theta)\sigma[q(s,\theta)]$$

$\sigma(q)$ is the scattering cross section of the interface and is taken as a lorentzian function centered at $q = 12.5nm^{-1}$ and with a FWHM equal to $0.002nm^{-1}$.

The incident intensity of the incident beam at the point $P(\theta, s)$ of the footprint:

$$I_0 = \begin{cases} e^{-\left(\frac{x}{w_x}\right)^2} e^{-\left(\frac{s\sin(\theta)}{w_y}\right)^2} & si\ x_P \in \left[-\frac{E}{2}, \frac{E}{2}\right] \\ 0 & si\ x_P \notin \left[-\frac{E}{2}, \frac{E}{2}\right] \end{cases}$$

<u>Determination of the longitudinal coordinate $x_P$ :</u>

In the $NH'M$ triangle, we can write :

$$\tan\theta = \frac{H'M}{H'N} = \frac{L\sin 2\theta_0 + u\cos 2\theta_0}{H'N}$$

The longitudinal coordinates of the $N$ point is given by:

$$\begin{aligned} x_N = OH - HN \quad &= L\cos 2\theta_0 - HH' - H'N \\ &= L\cos 2\theta_0 - u\sin 2\theta_0 - \frac{L\sin 2\theta_0 + u\cos 2\theta_0}{\tan\theta} \end{aligned}$$

Finally, the distance between the point $N$ and the projection of the point $P$ on the $x$-axis (incident beam axis) is $s\cos\theta$ and thus the longitudinal coordinate of the point $P$ is :

$$\begin{aligned} x_P \quad &= ON + NP \\ &= L\cos 2\theta_0 - u\sin 2\theta_0 - \frac{L\sin 2\theta_0 + u\cos 2\theta_0}{\tan\theta} + s\cos\theta \end{aligned}$$

<u>Determination of the scattering wave vector $\mathbf{q}(s,\theta)$ :</u>

We consider that the focalisation of the incident beam creates a distribution of angles for the incident scattering wave vector $\mathbf{k}_i$ between 0 and $\alpha_0$. The y-coordinate of the $P(s,\theta)$ point is $y_P = s\sin\theta$. Thus the angle between $\mathbf{k}_i$ and the direction of the incident beam is at point $P(s,\theta)$:

$$\alpha(s,\theta) = \alpha_0 \frac{y_P}{w_y/2} = 2\alpha_0 \frac{s}{w_y} \sin\theta .$$

Thus the coordinates of the wave vector of the incident beam at point $P(s,\theta)$ are

$$\mathbf{k}_i = \frac{2\pi}{\lambda}\begin{pmatrix} \cos\alpha \\ -\sin\alpha \end{pmatrix}$$

The coordinates of the wave vector of the scattered beam with the $\theta$ angle are :

$$\mathbf{k}_d = \frac{2\pi}{\lambda}\begin{pmatrix} \cos\theta \\ \sin\theta \end{pmatrix}$$

Finally, the scattering wave vector transfer is :

$$\mathbf{q} = \frac{2\pi}{\lambda}\begin{pmatrix} \cos\theta - \cos\alpha \\ \sin\theta + \sin\alpha \end{pmatrix}$$

The curves of figure 10-C are calculated using a software written in Python[36] using the standard integration routines of the Scientic Python module[37]

**Definitions:**

$L = OO'$ : Sample to detector distance

$\ell = OF$ : Sample to vertical slit distance

$2\theta_0$ : in-plane angle between the detector arm and the axis of the incident beam

$g = F_1F_2$ : gap of the vertical slit

$M$ : a point (horizontal axis) of the detector

$u = O'M$ : horizontal coordinate of the $M$ point on the detector

v : vertical coordinate of the $M$ point on the detector

$D$ : horizontal size of the 2D detector

**Table 1:**

| | |
|---|---|
| Sample to detector distance $L$ | $900mm$ |
| Sample to slit distance $\ell$ | $250mm$ |
| Horizontal size of the detector $D$ | $92mm$ |
| Wavelength $\lambda$ | $0.1605nm$ |
| $2\theta_0$ | $0.32rad$ |
| Longitudinal size of the footprint $E$ | $75mm$ |
| Longitudinal FWHM of the intensity distribution | $4.5mm$ |
| Transverse FWHM of the intensity distribution | $2mm$ |
| Convergence of the incident beam $\alpha_0$ | $6mrad$ |

**Captions:**

Figure 1: Top and side (inset) view of the classical grazing incidence x-ray diffraction setup of the D41 beam line at LURE.

Figure 2: Top view of the grazing incidence x-ray diffraction setup using a two dimensional detector and a single vertical slit on the $2\theta$ arm of the classical setup of figure 1.

Figure 3: Illustration of the resolution angle at point $M(u)$ of the detector in the top view of the 2D detector's setup. Inset: Illustration of the different angles for the calculation of the resolution of the setup.

Figure 4: Evolution along the horizontal direction of the detector of the setup's resolution $\Delta Q_{xy}$ computed for $D = 92mm$, $L = 595mm$, $\ell = 215mm$ and $g = 0.7mm$ and $2\theta_0 = 17.735$.

Figure 5: A) Evolution with the slit's gap of the resolution $\Delta Q_{xy}$ of the setup computed for $D = 92mm$, $L = 595mm$, $\ell = 215mm$, $2\theta_0 = 17.735$. B) Evolution of the resolution with the sample to slit distance $\ell$ computed for $D = 92mm$, $L = 595mm$, $g = 0.7mm$, $2\theta_0 = 17.735$. C) Evolution of the resolution with the sample to detector distance $L$ computed for $D = 92mm$, $g = 0.7mm$, $\ell = 215mm$, $2\theta_0 = 17.735$.

Figure 6: Picture of the setup built at LURE on the D41 beam line using a two dimensional detector and an horizontal gap single slit.

Figure 7: Diffraction spectra of a DPPE monolayer spread at the air-water interface compressed at surface pressure ($40mN.m^{-1}$) recorded with the two dimensional setup (line) and the classical 1D-PSD setup (points). For the 2D detector setup, the sample to detector distance is $L = 595mm$, the sample to slit distance is $\ell = 215mm$, the size of the detector is $92 \times 92mm^2$, the $2\theta$ arm is placed at an angle $2\theta_0 = 20.245°$ and the gap of the vertical slit is $g = 1mm$.

Figure 8: Successive diffraction spectra in $Q_{xy} - Q_z$ contour plot representation of a DPPE monolayer upon compression recorded with the new 2D detector setup; A : $5mN.m^{-1}$, B : $10mN.m^{-1}$, C : $15mN.m^{-1}$, D : $20mN.m^{-1}$, E : $25mN.m^{-1}$, F : $30mN.m^{-1}$, G : $35mN.m^{-1}$, H : $40mN.m^{-1}$. The acquisition time for each spectrum is $5mn$. The sample to detector distance is $L = 595mm$, the sample to slit distance is $\ell = 215mm$, the size of the detector is $92 \times 92mm^2$, the $2\theta$ arm is placed at an angle $2\theta_0 = 20.245°$ and the gap of the vertical slit is $g = 1mm$.

Figure 9: Successive diffraction spectra in $Q_{xy} - Q_z$ contour plot representation of a DPPE monolayer upon compression recorded with the 1D detector setup; A : $3mN.m^{-1}$, B : $20mN.m^{-1}$, C : $30mN.m^{-1}$, D : $40mN.m^{-1}$. The acquisition time for each spectrum is $120\,mn$ (2 hours).

Figure 10: A) Diffraction spectrum of a $CF_3 - (CF_2)_{10} - COOH$ monolayer spread at the air water interface at room temperature, compressed at surface pressure $20mN.m^{-1}$ and recorded with the 2D detector setup. The acquisition time is $3mn$. The sample to detector distance is $L = 595mm$, the sample to slit distance is $\ell = 215mm$, the size of the detector is $92 \times 92mm^2$, the $2\theta$ arm is placed at an angle $2\theta_0 = 17.735°$ and the gap of the vertical slit is $g = 0.25mm$. The line is a gaussian fit of the data point which gives the parameter of the peak (position and width) given inside the inset.

B) $Q_{xy} - Q_z$ contour plot of the diffraction spectrum of Fig 10-A showing the untilted nature of the perfluorinated chains.

C) Full Width at Half Maximum (FWHM) of the measured diffraction peaks as a function of the slit's gap. The lines are the result of the $\Delta Q$ calculation of Appendix 2 (thin continous line), and the best result of the calculation of appendix 3 (continuous thick line). The values of $L$, $\ell$, and $2\theta_0$ are given in Table 1.

D) Full Width at Half Maximum (FWHM) of the measured diffraction peaks as a function of the slit's gap. The lines are obtained by variations of the parameters of calculation of appendix 3. The values of $L$, $\ell$, and $2\theta_0$ are given in Table 1.

Figure 11: Illustration of the resolution calculation of appendix 3.

Figure 1

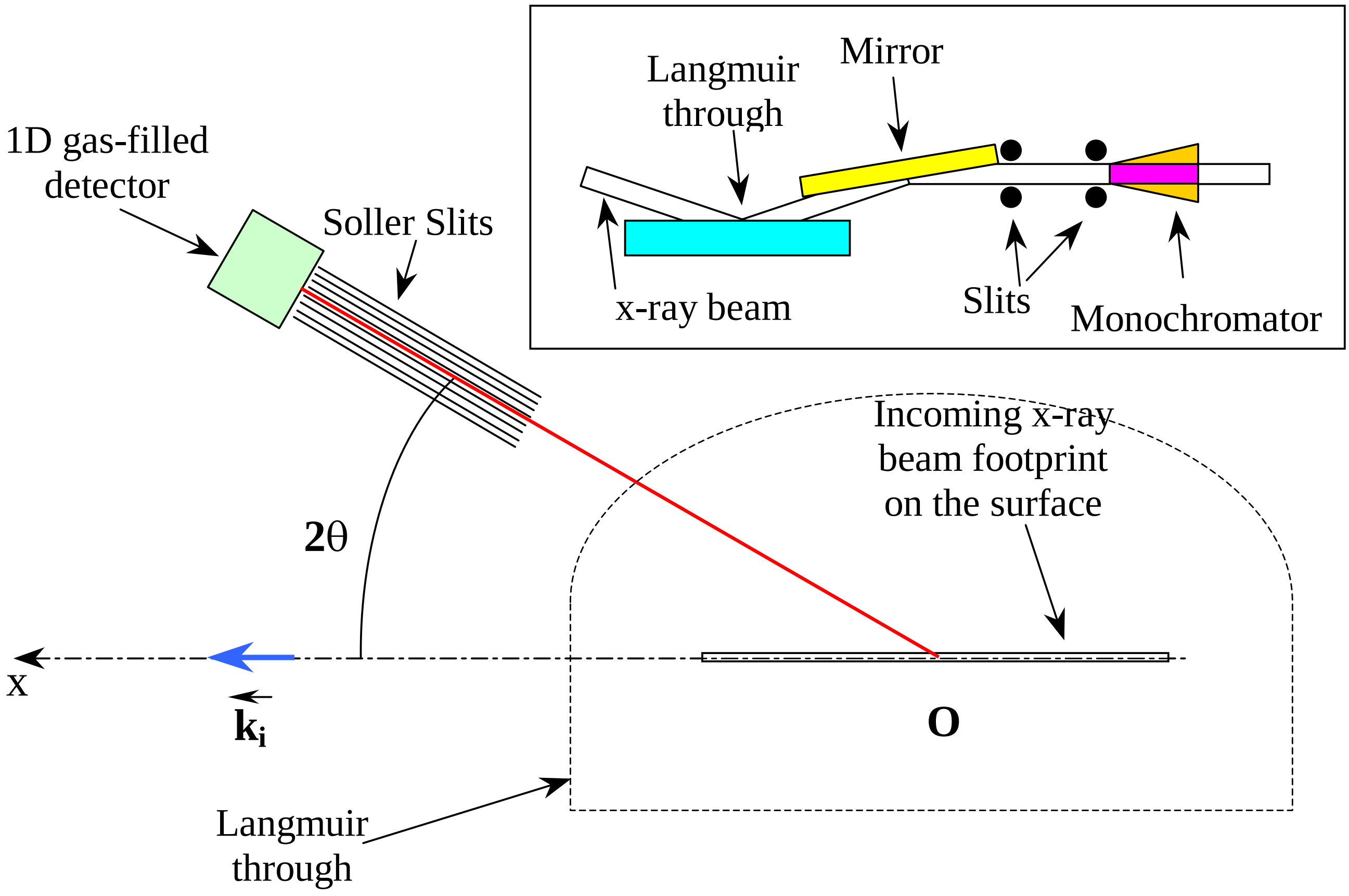

Figure 2

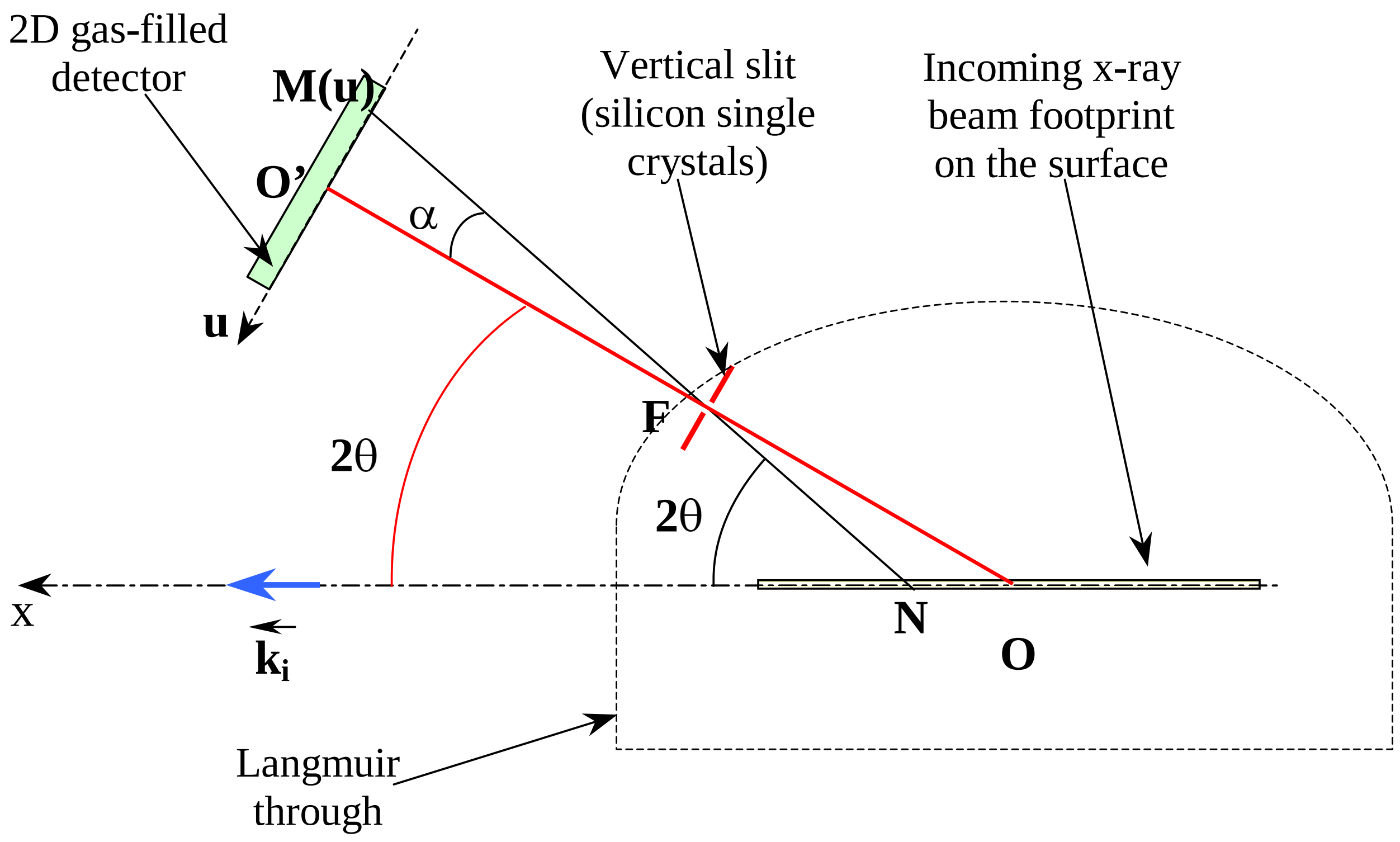

Figure 3

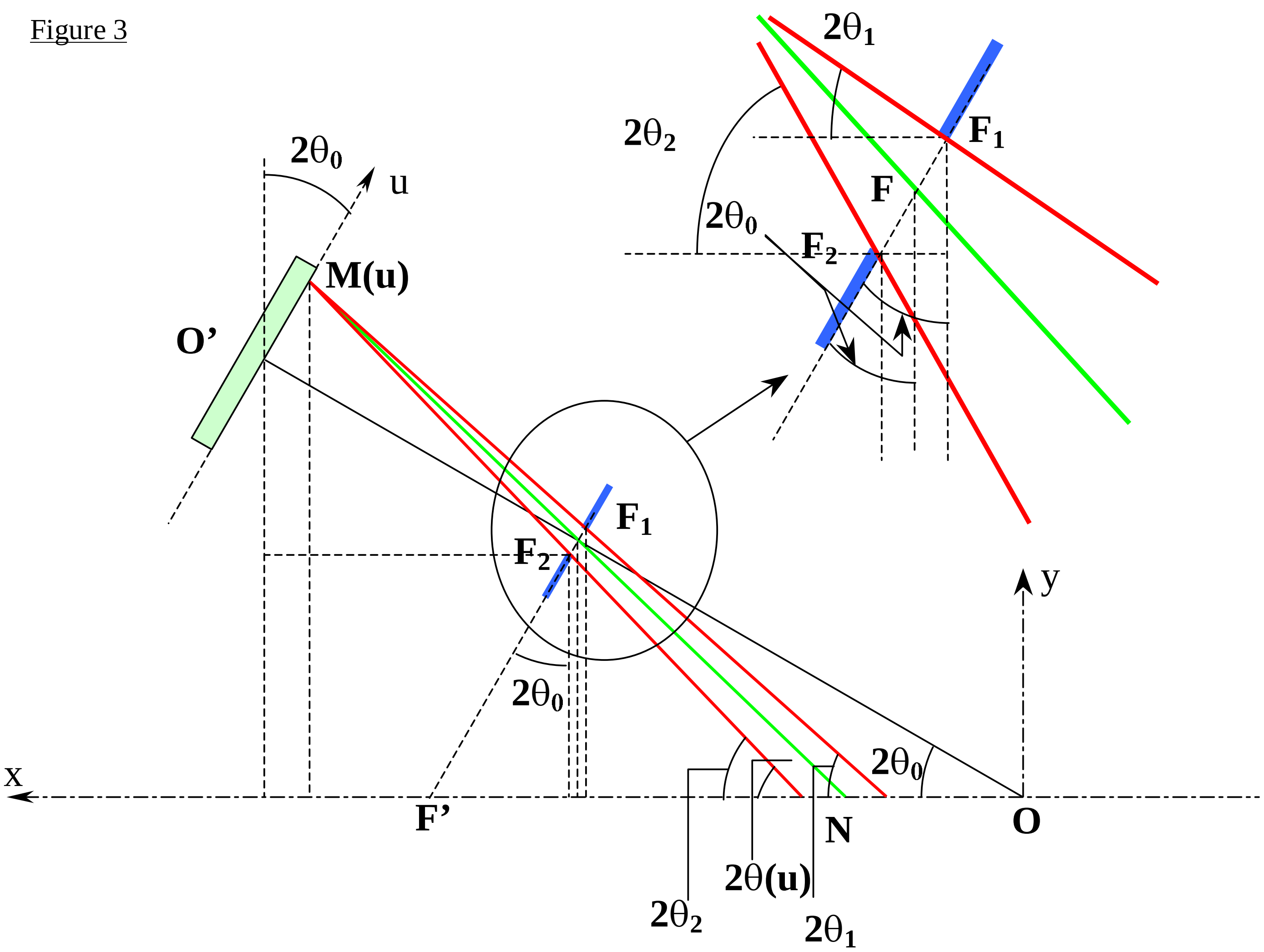

Figure 4

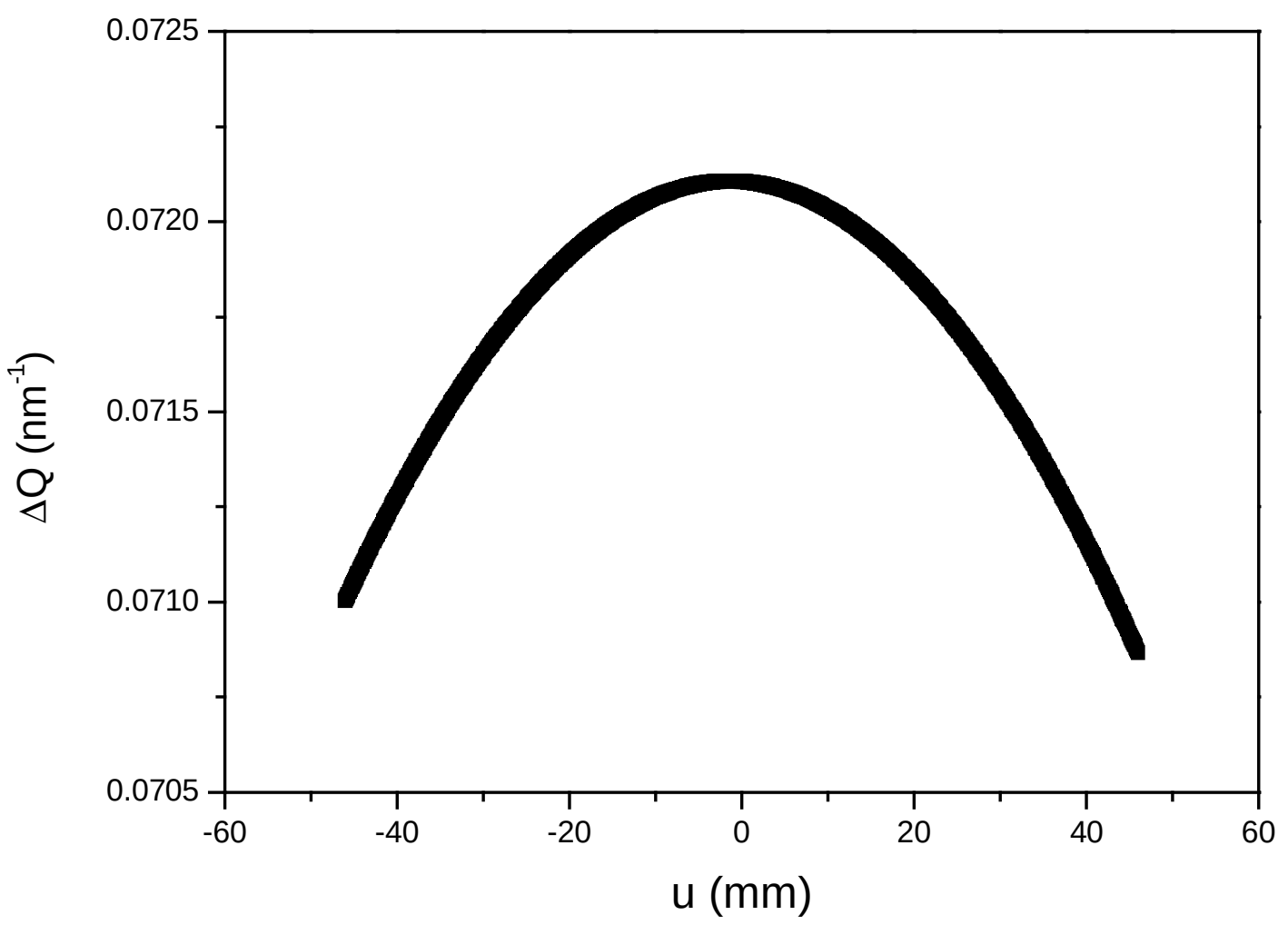

Figure 5:

A

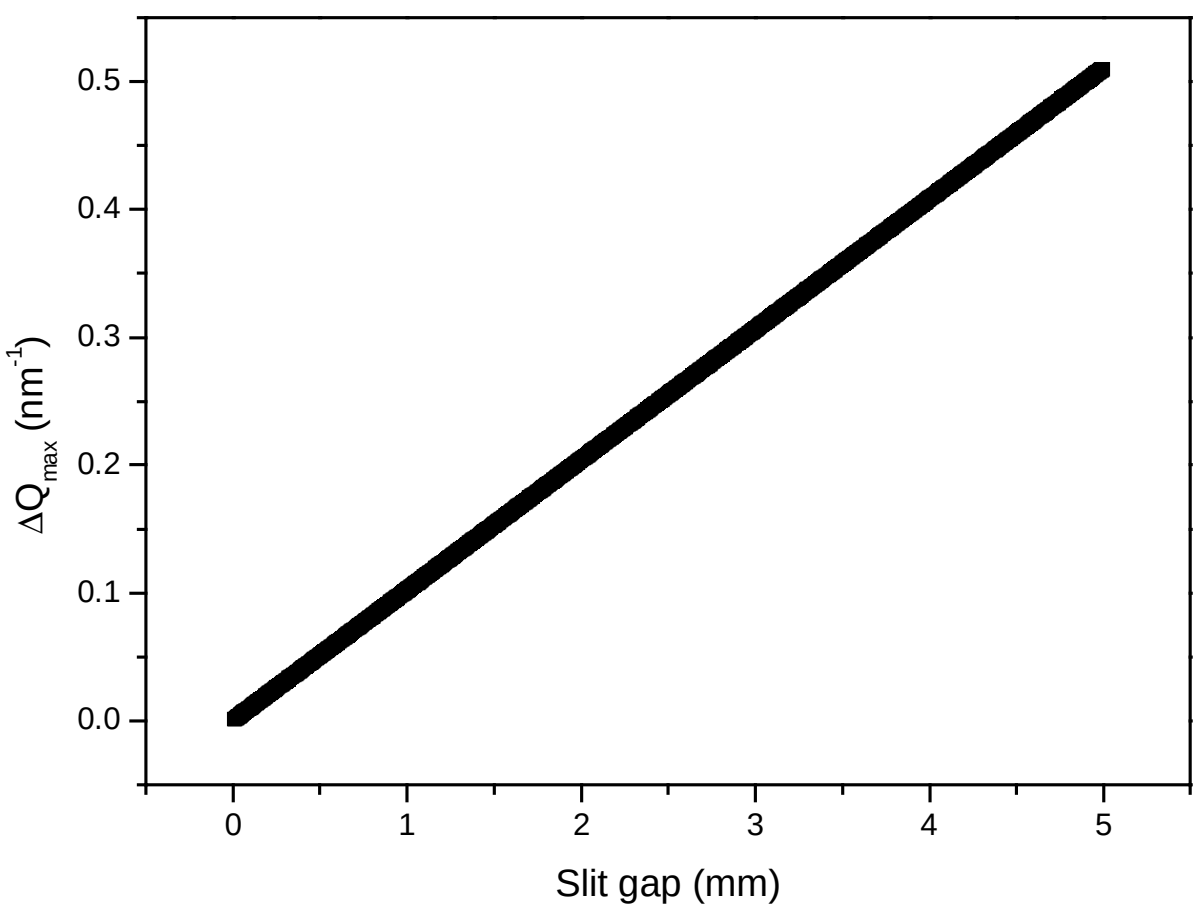

B

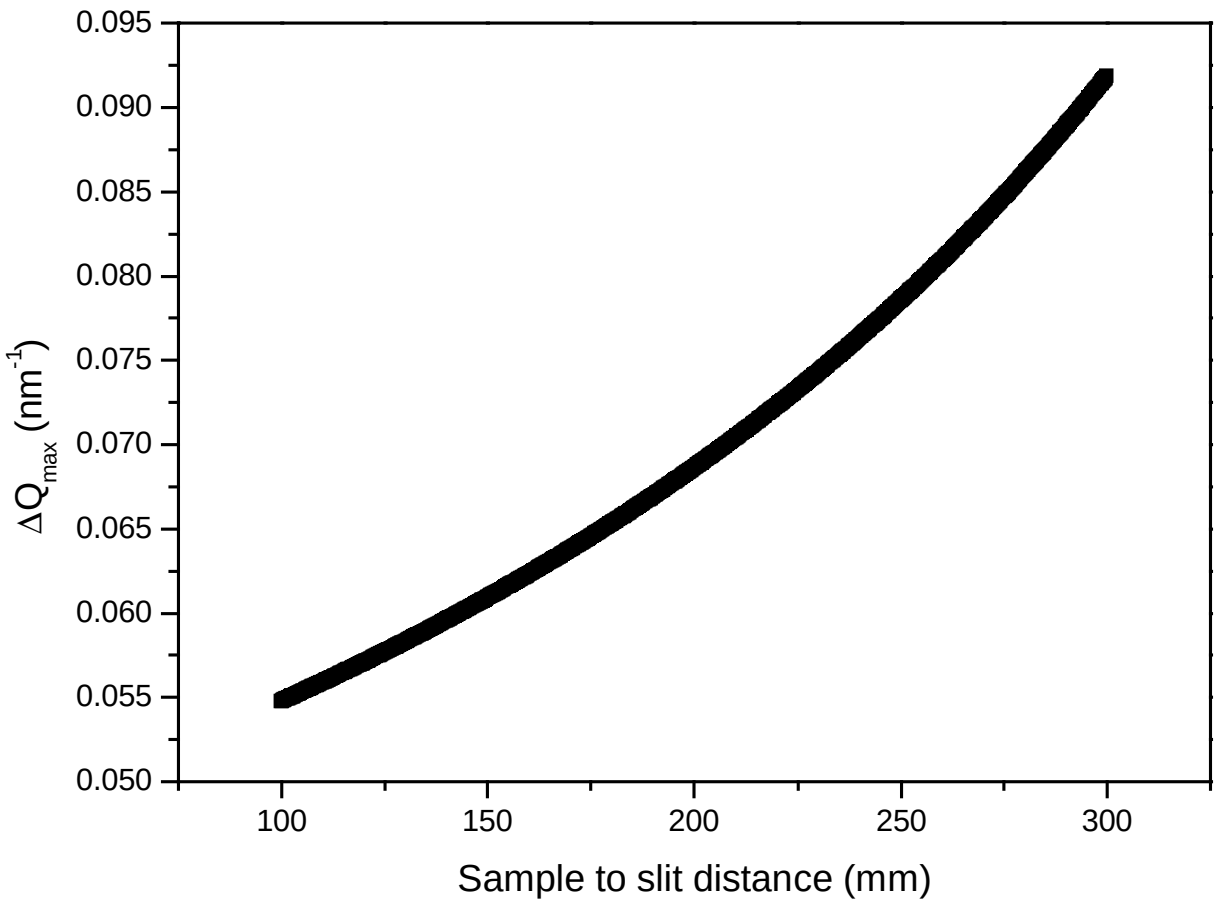

C

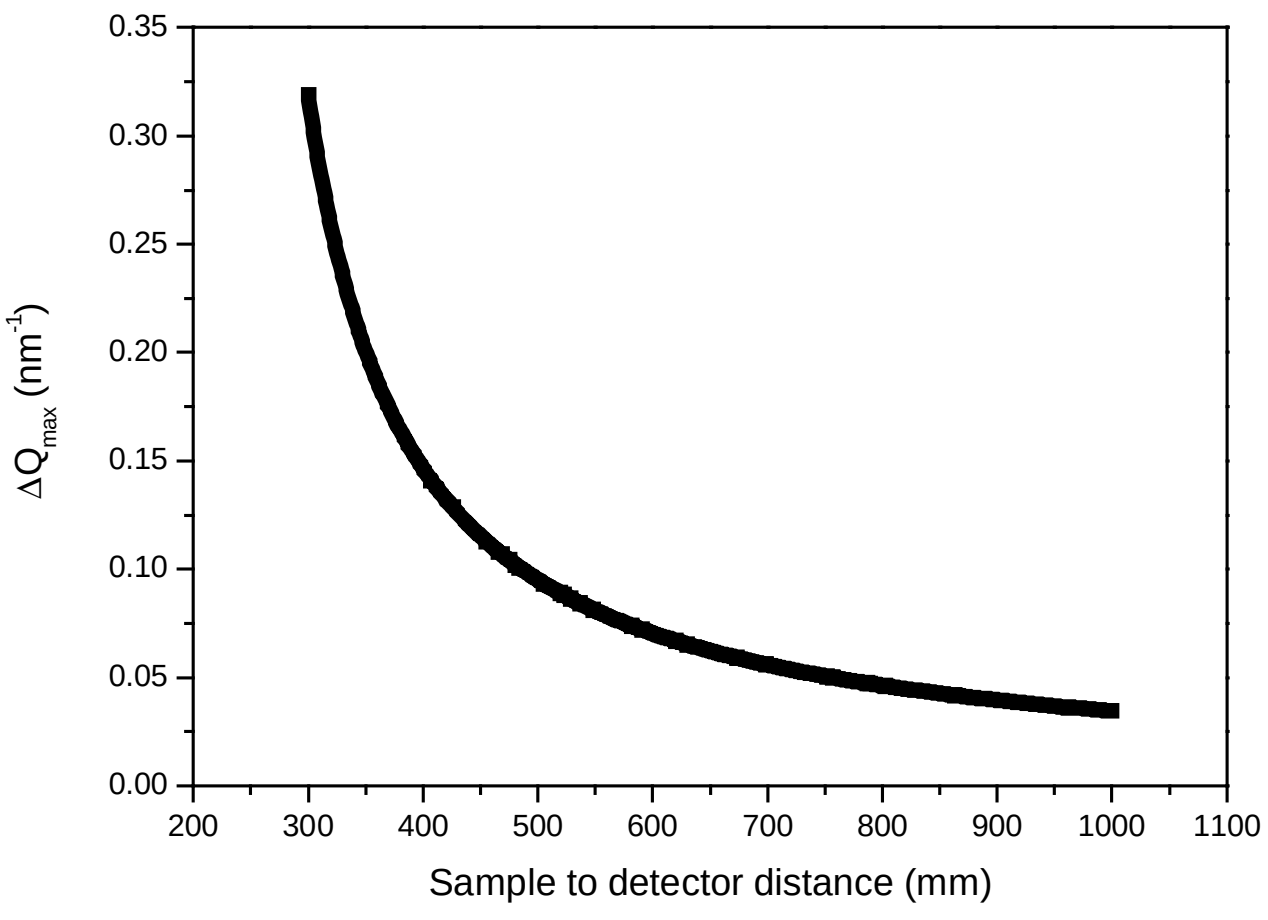

Figure 6 :

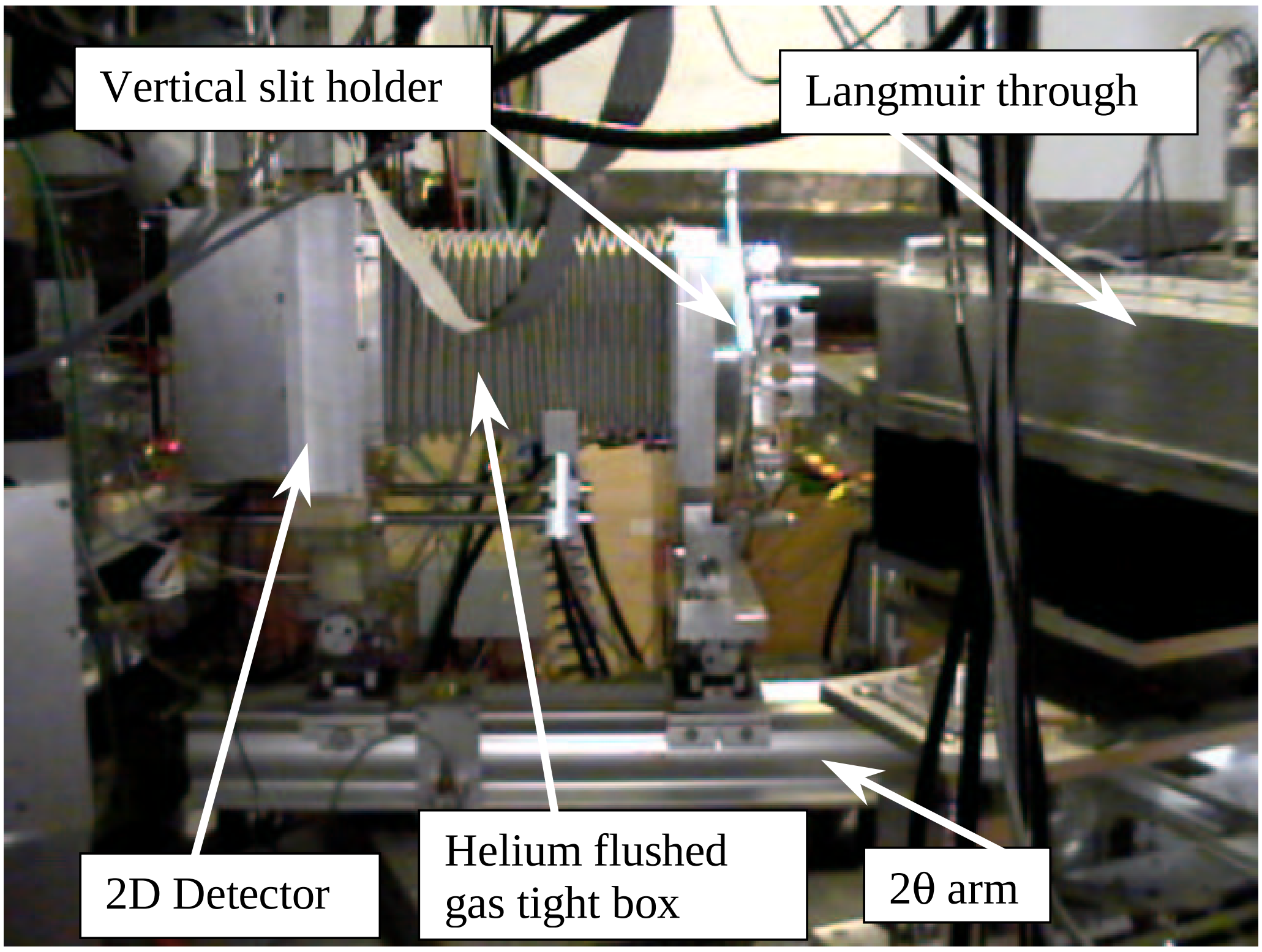

figure 7 :

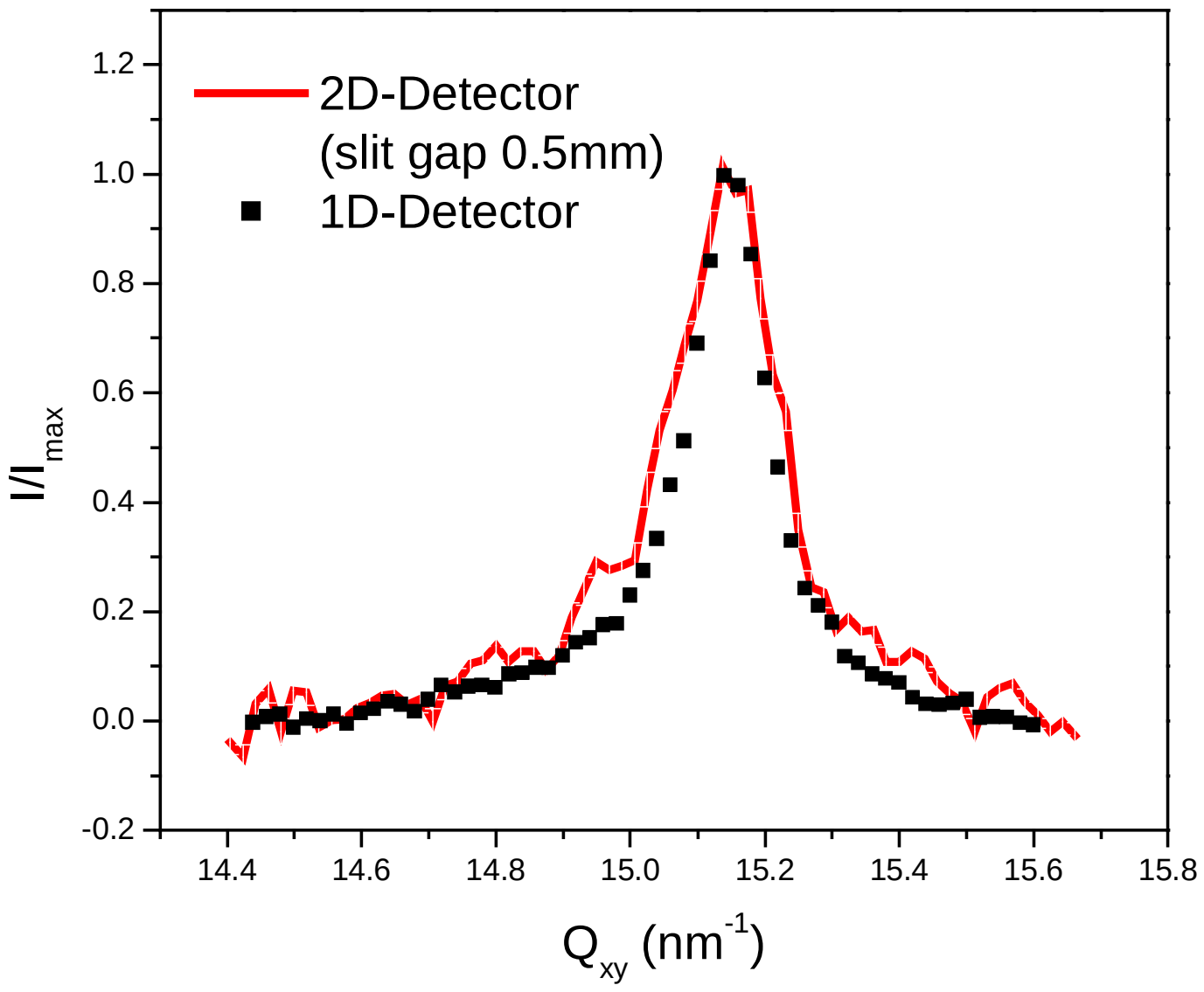

figure 8- :

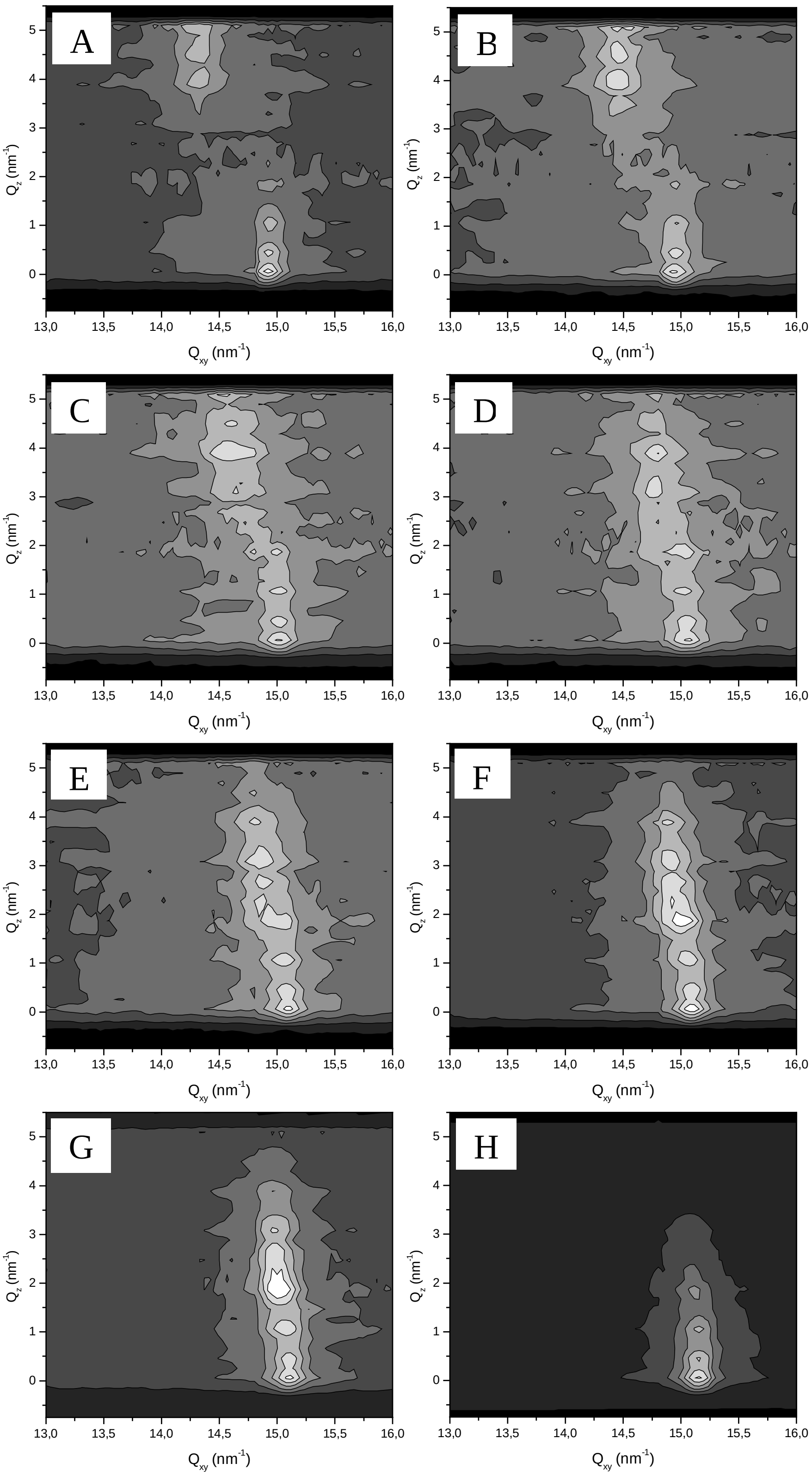

Figure 9

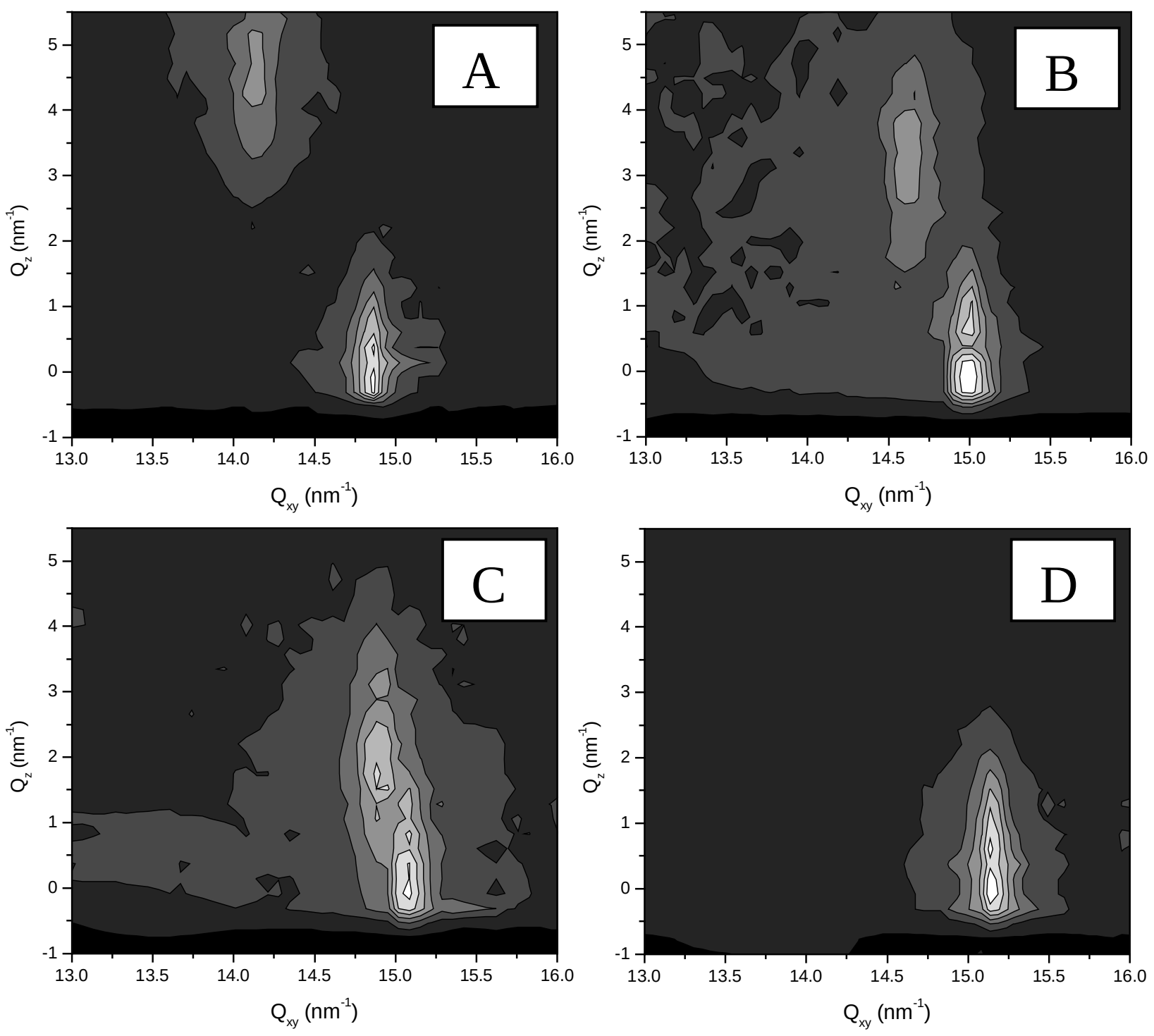

Figure 10

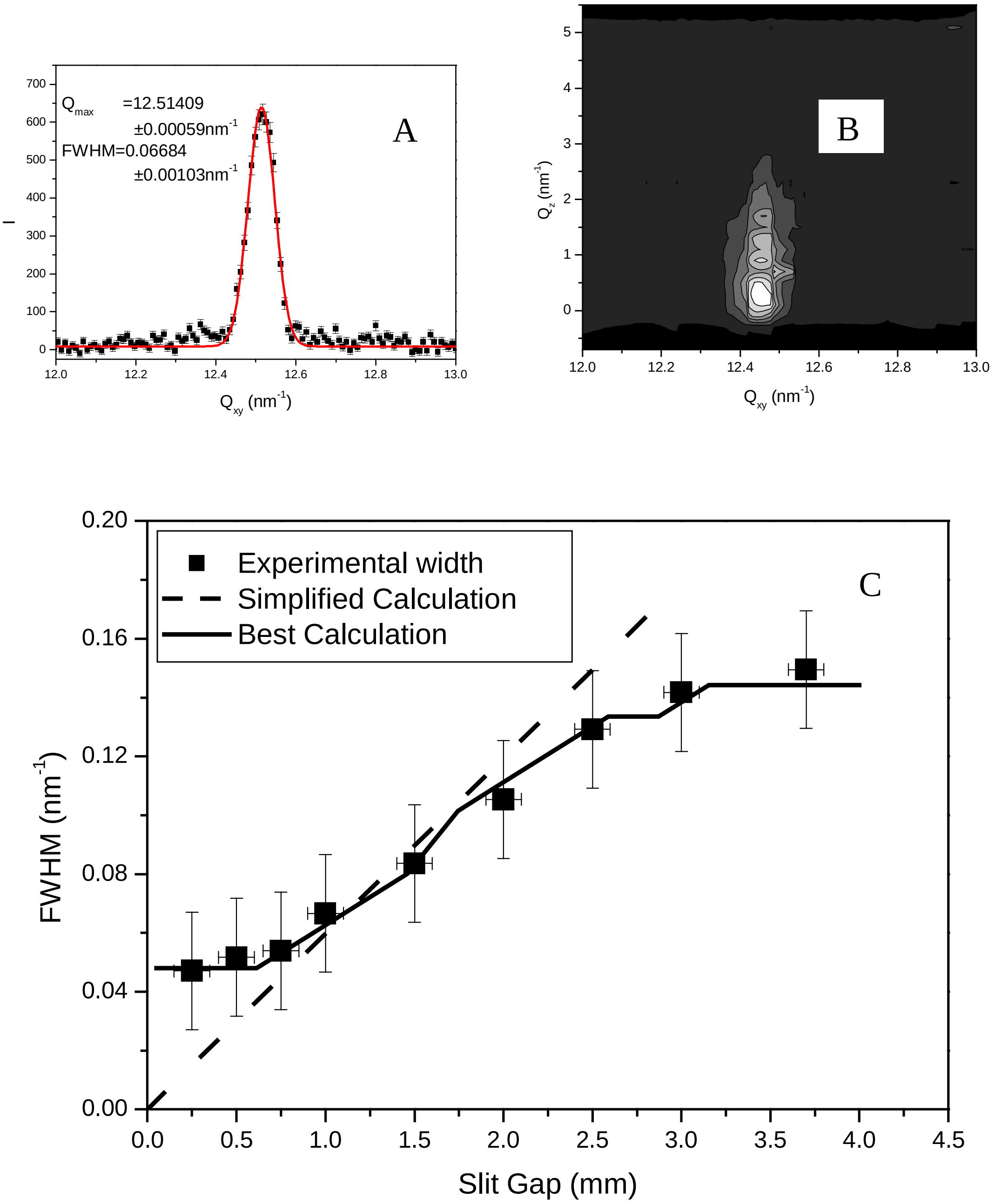

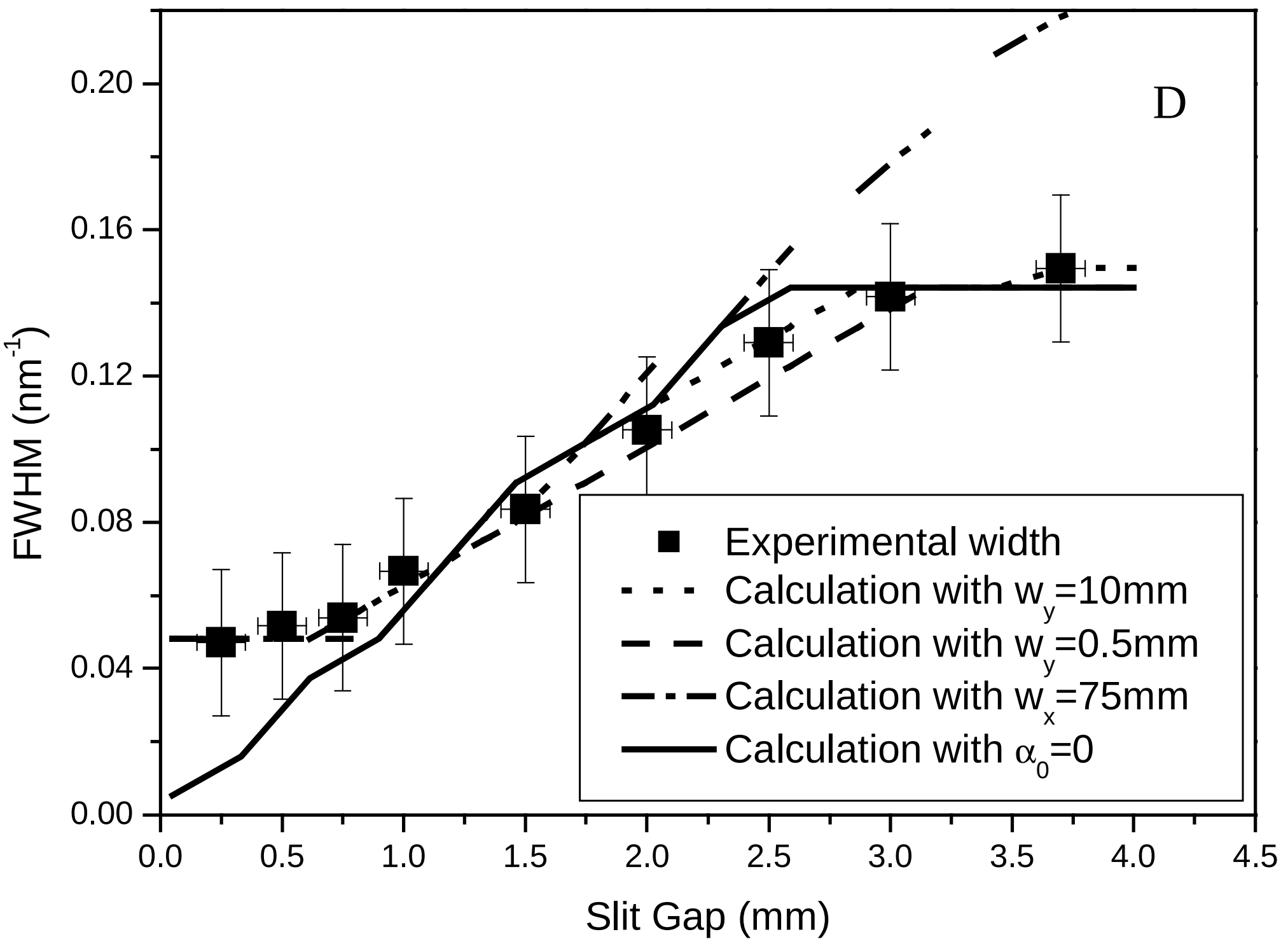

**Figure 11**

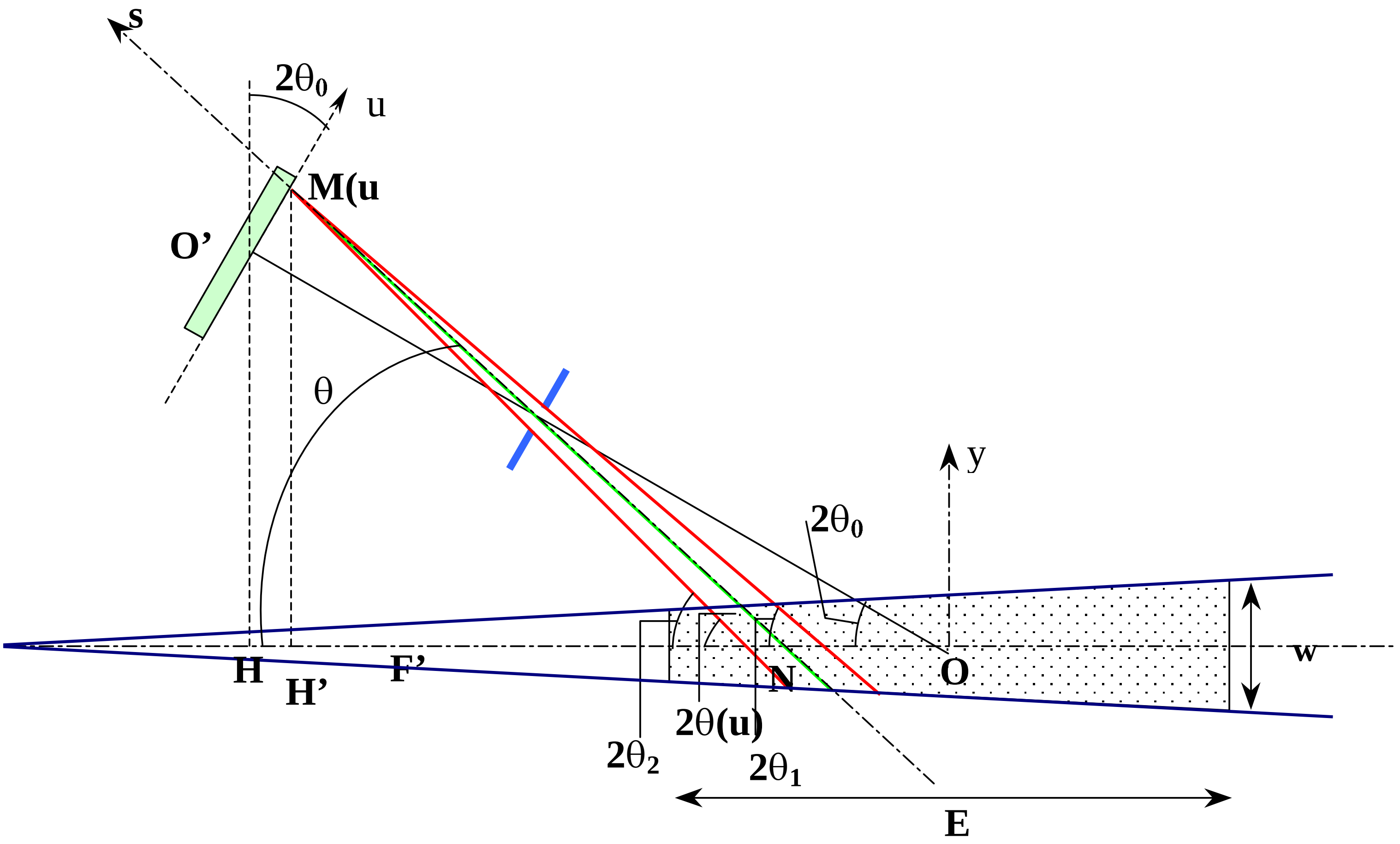